\documentclass[aps,pre, preprint,floatfix,showpacs,amsmath,amssymb]{revtex4}

\usepackage{graphicx}
\usepackage{dcolumn}
\usepackage{bm}
\usepackage{natbib}

\begin{document}

\title{Scattering a pulse from a chaotic cavity:  Transitioning from algebraic to exponential decay}

\author{James A. Hart}
\author{Thomas M. Antonsen Jr.}
\author{Edward Ott}
\author{Steven M. Anlage}

\affiliation{ Institute for Research in Electronics and Applied
Physics\\
University of Maryland\\
College Park, MD 20740 }

\date{\today}

\begin{abstract}
The ensemble averaged power scattered in and out of lossless
chaotic cavities decays as a power law in time for large times. In
the case of a pulse with a finite duration, the power scattered
from a single realization of a cavity closely tracks the power law
ensemble decay initially, but eventually transitions to an
exponential decay. In this paper, we explore the nature of this
transition in the case of coupling to a single port. We find that
for a given pulse shape, the properties of the transition are
universal if time is properly normalized. We define the crossover
time to be the time at which the deviations from the mean of the
reflected power in individual realizations become comparable to
the mean reflected power.  We demonstrate numerically that, for
randomly chosen cavity realizations and given pulse shapes, the
probability distribution function of reflected power depends only
on time, normalized to this crossover time.
\end{abstract}


\maketitle

\section{Introduction}

Waves and wave behavior are ubiquitous. Examples are acoustic
waves in matter, electromagnetic waves, and physical particles in
the quantum mechanical regime. Thus understanding wave behavior is
important in many different fields; systems which are radically
different physically can often be represented by the same
mathematics. The simplest model of wave behavior is the Helmholtz
equation,
\begin{equation}\label{eq:wave_equation}
    (\nabla^2+k^2)\phi=0,
\end{equation}
which typically must be supplemented with boundary conditions.
Equation~\eqref{eq:wave_equation} describes many physical
situations exactly (such as acoustic waves within a homogeneous,
linear, bulk medium or quantum particles in free space).
Inhomogeneous situations con often be modelled by
Eq.~\eqref{eq:wave_equation} with $k\rightarrow k(\vec{r})$ where
$k(\vec{r})$ is a function of position. If the system has loss or
gain, $k$ can be allowed to become complex. Driving terms can be
added to represent transducers or ports. In this paper, we focus
on scalar waves described by Eq.~\eqref{eq:wave_equation} with
constant $k$, but the results generalize well to many other wave
problems.

Unfortunately, for all but the simplest of geometries,
Eq.~\eqref{eq:wave_equation} is analytically intractable.  Thus
techniques, both numerical and theoretical, have been developed to
solve Eq.~\eqref{eq:wave_equation}.  These techniques and their
effectiveness vary depending on the regime and physical scenario
one wishes to study. In this paper, we limit ourself to the
semiclassical regime; i.e., the regime in which the wavelength of
the waves excited in the system is much shorter than the
scattering elements in the system.  In this limit, it is known
(via the correspondence principle from Quantum Mechanics) that the
resulting dynamics are closely related to the trajectories a
classical particle would take through the system. This analogy
applies even to purely classical waves, such as waves on the
surface of water where the role of classical particle dynamics is
now replaced by the dynamical evolution of ray trajectories. In
this paper, we consider only those systems in which the
corresponding classical dynamics is purely chaotic (i.e., all
classical trajectories which start infinitesimally far apart
diverge exponentially in time). In addition, we focus on the
scattering properties of such systems, assuming that the system of
interest is a closed cavity that couples to the outside world only
via well-defined localized channels.

The scattering properties of such wave systems have been well
studied, both experimentally
\cite{Weaver_acoustic,Ellegard_acoustic,Chinnery1,Chinnery2,Lindelof,Bluemel}
and theoretically
\cite{Porter,Alhassid_review,Guhr_WeidenMueller_MuellerGroeling_review,Dittes_Harney_Muller,darmstadt_group_algebraic,
Zozoulenko_and_Blomquist,Buttiker_alone}, in a wide variety of
contexts. Much of the the theory has focused on the frequency
domain, and sophisticated techniques exist to analyze and
characterize the scattering process. See
Refs.~\cite{Alhassid_review} and
\cite{Guhr_WeidenMueller_MuellerGroeling_review,Dittes_Harney_Muller,darmstadt_group_algebraic,
Zozoulenko_and_Blomquist,Buttiker_alone} and the references cited
therein. Similarly, the time domain response of typical wave
systems to a delta-function impulse has also been considered
\cite{Dittes_Harney_Muller,darmstadt_group_algebraic,
Zozoulenko_and_Blomquist,Buttiker_alone}, especially in
relationship to fidelity decay(for an overview of fidelity decay,
see the Ref.~\cite{Fidelity_review_article} and the references
therein). In this paper we consider an intermediate situation: we
excite the wave system, through an external port with a pulse
modulated sinusoidal signal, exciting a large but finite number of
modes.  The problem of scattering pulse-modulated sinusoidal waves
arises in a host of diagnostic situations, such as radar, sonar,
nuclear scattering, etc.  In what follows, for specificity, we
discuss our problem in the context of electromagnetic waves. For
simplicity, we consider only lossless two-dimensional microwave
cavities excited through a small antenna.  We emphasize that the
results we obtain can be generalized to higher-dimensional systems
and to quantum mechanical or other wave-chaotic systems(e.g.,
acoustic or elastic wave systems).

On a formal level, the time domain dynamics of such a system is
straightforward. The system is open and linear.  An incident pulse
with a small but finite width in the time domain excites a large
number of modes in the cavity, which then radiate their energy
back out through the port. Because the system is linear, the
reflected voltage can be expressed as a superposition of
contributions from modes of the open system. The chaotic dynamics
is expressed, not through the dynamics of the individual modes,
but rather in the eigenvalue statistics \cite{gutzwiller_book} and
the statistics of the coupling between the port and the cavity.

As showed in Sec.~\ref{sec:model}, the contribution from each mode
decays exponentially in time.  For short times compared with the
Heisenberg time (the inverse of the mean spacing of mode
frequencies), the resulting dynamics will be determined primarily
by the semiclassical dynamics within the cavity
\cite{Schomerus_early_time_decay}. However, for large times
compared with the Heisenberg time, the ensemble average of the
reflected power decreases as a power law in time
\cite{Dittes_Harney_Muller}. This is due to the fact that there is
a probability distribution of mode decay rates which extends to
zero decay rate, and for long times the average is dominated by
modes with very small decay rates. In the case of a single
realization of the chaotic cavity, the incident pulse excites a
large number of modes with very similar amplitudes, and
consequently the reflected power initially behaves as though the
sum of modes were an ensemble average, and the total power decays
as a power law. We call this behavior self-averaging. In a single
specific realization, however, there are only a finite number of
modes excited. Eventually the slowest-decaying mode in the
realization will be much larger than the other modes, and the sum
will be dominated by this slowest mode, which decays
exponentially.  Thus for extremely long times we expect that the
reflected power for any single realization will fall
exponentially, eventually becoming much smaller than the ensemble
average.

\begin{figure*}
\includegraphics{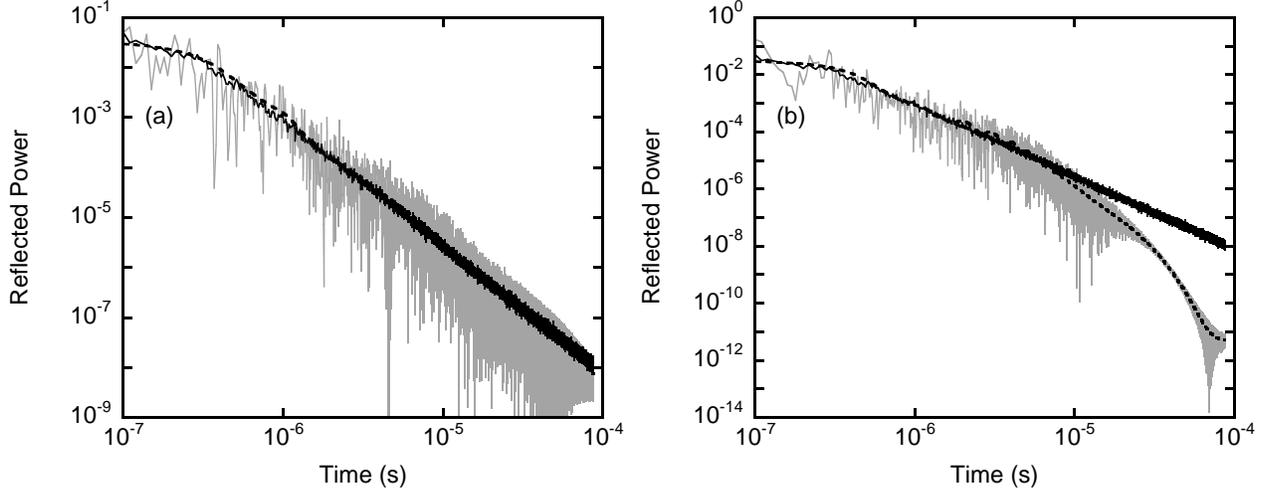}
\caption{\label{fig:single_realizations} Using the Random Coupling
Model (RCM), we created a program capable of simulating the
time-domain response of an individual chaotic cavity to a pulse
injected into the cavity through a small antenna.  By repeatedly
creating individual cavities using the RCM, we created an ensemble
of such cavities. The gray lines represent the power reflected
back into the cavity from two single realizations of the chaotic
cavity. The dark solid line represents the reflected power
averaged over 50 realizations of the chaotic cavity. The dashed
line represents the time-averaged power for the single
realization.  Figure (a) represents a cavity where self-averaging
persists throughout the entire simulation, but figure (b) is
dominated by solitary modes after about $10^{-5}$ seconds.}
\end{figure*}

To test this hypothesis, we have created a program that models the
time-domain behavior of generic chaotic systems. It does this by
first generating the spectrum and coupling constants of a cavity
using the previously published \cite{Henry_paper_one_port} Random
Coupling Model (RCM) and then integrating the evolution equations
for fields in the cavity, which are modelled in the RCM as a set
of driven, damped coupled harmonic oscillators. Single
realizations of the power reflected from these cavities, as well
as the ensemble average of 50 different cavities, are shown in
Fig.~\ref{fig:single_realizations}, where we show two very
different realizations: one
(Fig.~\ref{fig:single_realizations}(a)) in which the
self-averaging persists throughout the length of the time shown
and one (Fig.~\ref{fig:single_realizations}(b)) in which
self-averaging occurs early, but becomes dominated by solitary
slowly decaying modes before the conclusion of the numerical
simulation.

Our goal in this paper is to quantitatively describe the
transition from self-averaging to exponential decay.  In
particular, we wish to predict the time-scale needed to see this
transition. In Sec.~\ref{sec:model}, we describe the time-domain
model we use for our analysis.  In Sec.~\ref{sec:Finding_P_gamma},
we find the probability distribution function of the decay rates
of the open-cavity modes (for the slowest decaying modes in the
cavity) as a function of the cavity's port reflection coefficient.
In Sec.~\ref{sec:bar_P_stats} we find the average, standard
deviation and (indirectly) the higher-order moments of the
reflected power as a function of time, and use these moments to
derive a normalized time which, along with the power spectrum of
the incident pulse, is all that is needed to obtain a
characterization of the transition from self-averaging to
exponential decay.  In Sec.~\ref{sec:numerics}, we evaluate the
theory from Sec.~\ref{sec:bar_P_stats} by numerically finding the
number of modes which fall below certain fractions of the average,
and we compare the theory with simulation results.

\section{Model}\label{sec:model}

We base our model system on that used in previous work
\cite{Henry_paper_one_port}; specifically a quasi-two-dimensional,
electromagnetic cavity defined by two conducting plates of area
$A$ separated by a distance $h$ which are electrically connected
along their perimeters by a conducting side-wall.  The cavity is
excited by an antenna that induces currents in the plates.  The
wave equation for this system is
\begin{equation}\label{eq:startEq}
\frac{1}{c^2}\frac{\partial^2}{\partial t^2}V_T-\nabla^2V_T=h\mu
u\frac{\partial I}{\partial t},
\end{equation}
where $c=(\epsilon\mu)^{-1/2}$ is the speed of propagation of
waves in the uniform medium inside the cavity, $\epsilon$ and
$\mu$ are the permittivity and permeability of this
(non-dispersive) medium, $V_{T}(x,y)$ is the voltage difference
between the plates, an antenna is modelled through the function
$u(x,y)$ which gives the profile of current flowing in the antenna
between the surfaces ($\int\int dx\,dy\,u(x,y)=1$), and $I(t)$ is
the time-dependent current driving the antenna. Further, as the
side walls of the cavity are conducting, $V_{T}=0$ along the
perimeter of the cavity.  A voltage $V(t)$ is induced at the
terminals of the model antenna which is given in terms of the
antenna profile $u$ and $V_{T}$
\begin{equation}\label{eq:V_def}
    V=\int dx\,dy\,uV_{T}.
\end{equation}

The antenna is excited by an incident voltage pulse
$V_{\textrm{inc}}(t)$ arriving along a transmission line of
characteristic impedance $Z_0$.  The incident wave excites the
cavity and produces a reflected wave pulse $V_{\textrm{ref}}(t)$
travelling away from the cavity in the transmission line. At the
junction between the transmission line and the cavity the voltages
and currents at the antenna and on the transmission line match,
\begin{eqnarray}
  V(t) &=& V_{\textrm{inc}}(t)+V_{\textrm{ref}}(t)\label{eq:V_sum}, \\
  I(t) &=& Z_0^{-1}[V_{\textrm{inc}}(t)-V_{\textrm{ref}}(t)].
  \label{eq:I_V_diff}
\end{eqnarray}

We now introduce Fourier transforms with transform frequency
$\omega$ such that each time-dependent variable is represented in
the following way,
\begin{equation}\label{eq:V_omega_def}
    V_{T}(x,y,t)=\int \frac{d\omega}{2\pi}e^{j\omega t}
    \bar{V}_{T}(x,y,\omega).
\end{equation}
The transformed field within the cavity is then represented as a
superposition of the orthonormal modes of the closed cavity,
\begin{equation}\label{eq:v_omega_sum}
    \bar{V}_{T}(x,y,\omega)=\sum_n c_{n}(\omega)\phi_{n}(x,y).
\end{equation}
where $(\nabla_{x,y}^2+k_n^2)\phi_n=0$, and $\phi_n=0$ on the
cavity side walls.

Solving the transformed wave equation gives the amplitudes
$c_n(\omega)$ which can then be inserted in Eq.~\eqref{eq:V_def}
to find the transformed voltage,
\begin{equation}\label{eq:impedance_used}
    \bar{V}(\omega)=\bar{I}(\omega)Z_{e}(\omega),
\end{equation}
where
\begin{equation}\label{eq:impedance_as_sum}
    Z_{e}(\omega)=-j\sqrt{\frac{\mu}{\epsilon}}\sum_{n}\frac{k
    h}{k^2-k_n^2}\left[\int dx\,dy\,u\phi_n\right]^2
\end{equation}
is the (exact) cavity impedance.  Here $k_n^2$ are the eigenvalues
of the closed cavity and $k=\omega/c$.

In Ref.~\cite[Eq.~14]{Henry_paper_one_port}, it was shown that, if
one assumed for the purpose of evaluating
Eq.~\eqref{eq:impedance_as_sum} that the eigenfunctions behave as
if they were a superposition of random plane waves, the overlap
between the eigenfunctions and antenna current profile could be
expressed in terms of the radiation resistance of the antenna,
\begin{equation}\label{eq:R_rad_def}
    R_{rad}(k)=\frac{k
    h}{4}\sqrt{\frac{\mu}{\epsilon}}\int\frac{d\theta}{2\pi}|\bar{u}(\vec{k})|^2,
\end{equation}
where $\bar{u}(\vec{k})$ is the spatial Fourier transform of the
profile function $u(x,y)$, and the integral is over the angle
$\theta$ of the vector $\vec{k}$.

Here $R_{Rad}=\textrm{Re}[Z_{Rad}]$ where $Z_{Rad}$, the radiation
impedance, is the impedance $\bar{V}(\omega)/\bar{I}(\omega)$ that
would apply if the cavity side walls were moved to infinity and
outward propagating radiation conditions were imposed.

With this random plane wave assumption, the exact impedance
$Z_{e}$ in Eq.~\eqref{eq:impedance_as_sum} was replaced by a
statistical model impedance,
\begin{equation}\label{eq:statistical_impedance}
    Z(\omega)=-\frac{j}{\pi}\sum_n\frac{k\Delta
    w_n^2}{k^2-k_n^2}\frac{R_{Rad}(k_n)}{k_n},
\end{equation}
where $w_n$ are zero mean, unit variance, independent Gaussian
random variables.  It was further assumed in
Ref.~\cite{Henry_paper_one_port} that the eigenvalues $k_n^2$ have
the statistical properties of eigenvalues of a Gaussian Orthogonal
Ensemble (GOE) random matrix with mean spacing given by Weyl's
formula,
\begin{equation}\label{eq:Weyls_formula}
\langle k_{n+1}^2-k_n^2 \rangle_n\equiv\Delta=4\pi/A.
\end{equation}

We now use the relationship (Eq.~\eqref{eq:impedance_used})
between the voltage $\bar{V}(\omega)$ and current
$\bar{I}(\omega)$ along with the transformed version of Eqs.
(\ref{eq:V_sum}) and (\ref{eq:I_V_diff}) to find the transform of
the reflected voltage pulse,
\begin{equation}\label{eq:basic_scattering}
    \bar{V}_{\textrm{ref}}(\omega)=\rho(\omega)\bar{V}_{\textrm{inc}}(\omega),
\end{equation}
where the reflection coefficient $\rho(\omega)$ is given by
\begin{equation}\label{eq:rho_def}
    \rho(\omega)=\frac{Z(\omega)-Z_0}{Z(\omega)+Z_0}.
\end{equation}
Although the derivation above has focused on the electromagnetic
case, the expression Eq.~\eqref{eq:rho_def} describes the
reflection of a wide variety of waves when they hit an interface,
viz., electromagnetic, acoustic, quantum mechanical, etc. The
connection becomes closer when one considers, as we will, incident
pulses whose transformed bandwidth $\omega_B$ is narrow enough
that the radiation resistance and mean frequency spacing can be
considered constant over the range of excited frequencies.

The time-dependence of the reflected pulse can be found by using
the inverse Fourier transformation,
\begin{equation}\label{eq:V_ref_time_domain}
    V_{\textrm{ref}}(t)=\int
    \frac{d\omega}{2\pi}\rho(\omega)\bar{V}_{\textrm{inc}}(\omega)
    e^{j\omega t}.
\end{equation}
The long-term behavior of the reflected pulse is governed by the
poles of $\rho(\omega)$ (denoted $\omega_k$), which satisfy
\begin{equation}\label{eq:Z_sum_zero}
    Z_0+Z(\omega_k)=0.
\end{equation}
The complex frequencies $\omega_k$ have positive imaginary parts
as they correspond to decaying modes.  We can approximate the long
time dependence of the reflected pulse by pushing the inversion
contour in Eq.~\eqref{eq:V_ref_time_domain} up into the upper half
of the $\omega$-plane and deforming it around each pole
\begin{equation}\label{eq:V_ref_time_domain_sum}
    V_{\textrm{ref}}(t)=-2j\sum_{k}\frac{Z_0 }
    {Z'(\omega_k)}\bar{V}_{\textrm{inc}}(\omega_k)e^{j\omega_k t},
\end{equation}
where $Z'(\omega_k)=dZ/d\omega|_{\omega=\omega_k}$. Thus, the long
time behavior of $V_{\textrm{ref}}(t)$ is determined by the
properties of eigenfrequencies $\omega_k$ of the open system.
These eigenfrequencies have real values whose average spacing is
denoted by $\Delta\omega$.  In principle, $\Delta\omega$ can vary
as a function of mode number.  If we assume that the incident
pulse has a spectrum centered at a carrier frequency $\omega_0$,
with a bandwidth $\omega_B\ll \omega_0$ we can relate
$\Delta\omega$ to the mean spacing $\Delta$ of $k_n^2$ values
\begin{equation}\label{eq:Delta_omega_result}
    \Delta\omega=\frac{c^2\Delta}{2\omega_0}.
\end{equation}
The inverse of this quantity can be identified with what is known
as the Heisenberg time in the Quantum Chaos community.

Each mode has a decay rate $\gamma_k=\textrm{Im}(\omega_k)$ which
varies from mode to mode.  We denote the probability density
function of these decay rates by $P_{\gamma}(\gamma)$. Considering
the number of excited modes to be effectively finite, since each
mode decays exponentially, the long time behavior of the reflected
signal is dominated by modes with the smallest values of
$\gamma_k$. From Eq.~\eqref{eq:Z_sum_zero}, along with the
expression for $Z(\omega)$ in
Eq.~\eqref{eq:statistical_impedance}, it can be seen that these
weakly coupled modes will have particularly small $w_n$ and thus
$\textrm{Re}(\omega_k)\simeq k_n c$. Given this observation, we
can approximate the complex mode frequencies $\omega_n$ by solving
for the poles in the weak coupling approximation. Specifically, in
Eq.~\eqref{eq:statistical_impedance}, our expression for the
impedance, we separate the term with $\omega_n\simeq k_n c$ from
the others,
\begin{equation}\label{eq:impedance_separate_sum}
    Z(\omega_n)=j X_n -j \frac{R_{Rad}(\omega_0) \Delta\omega w_n^2}
    {\pi(\omega_n-k_n c)},
\end{equation}
where we have changed our indexing labels from $k$ to $n$ (because
every $k_n$ has a corresponding $\omega_n$), and
\begin{equation}\label{eq:X_n_def}
    X_n=-\frac{1}{\pi}\sum_{n'\neq n}\frac{k_n
    w_{n'}^2\Delta}{k_n^2-k_{n'}^2}\frac{R_{Rad}(k_{n'})}{k_{n'}}.
\end{equation}
Thus, we can solve Eq. \eqref{eq:Z_sum_zero} approximately for the
complex mode frequencies,
\begin{equation}\label{eq:complex_mode_solution}
    \frac{\omega_n-k_n c}{\Delta\omega}=j w_n^2 \frac{R_{Rad}}{\pi(Z_0+j
    X_n)}.
\end{equation}
From this we obtain an expression for the decay rate,
\begin{equation}\label{eq:gamma_n_solution}
    \gamma_n=\Delta\omega w_n^2 \frac{R_{Rad}
    Z_0}{\pi(Z_0^2+X_n^2)}.
\end{equation}
The reactance $X_n$, like the impedance $Z$ is a statistical
quantity.  It has an average value to which all the terms in
Eq.~\eqref{eq:X_n_def} contribute, and which can be calculated by
replacing the sum by an integral \cite{Henry_paper_one_port},
\begin{equation}\label{eq:X_sum_integral}
    \langle X_n\rangle = X_{Rad}=-\frac{1}{\pi}P\left\{\int_{0}^{\infty}
    dk_{n'}^2\,\frac{k_n}{k_{n'}}\frac{R_{Rad}(k_{n'})}{k_n^2-k_{n'}^2}
    \right\}.
\end{equation}
where the symbol $P$ indicates that principal value definition of
the the integral is to be taken.  This average value is the
radiation reactance of the antenna. The reactance $X_n$ has a
fluctuating part which scales as the radiation resistance and is
due primarily to terms in the sum where $n$ and $n'$ are not too
different,
\begin{equation}\label{eq:X_n_decomposed}
    X_n=X_{Rad}+R_{Rad}\xi_n.
\end{equation}
The quantity $\xi_n$ has a universal distribution which we will
investigate in depth later.

Using Eqs.~(\ref{eq:impedance_separate_sum}) and
(\ref{eq:complex_mode_solution}) we may evaluate $Z'(\omega_n)$ in
the denominator of Eq.~\eqref{eq:V_ref_time_domain_sum}.  The
result for the reflected signal is
\begin{equation}\label{eq:V_ref_in_t}
    V_{\textrm{ref}}(t)=-2\sum_n \frac{Z_0
    R_{Rad}}{(Z_0+jX_n)^2}w_n^2e^{j\omega_n
    t}\Delta\omega\bar{V}_{\textrm{inc}}(\omega_n).
\end{equation}

Taking the magnitude of this, we obtain the reflected power,
\begin{equation}\label{eq:P_ref_split_terms}
    P_{\textrm{ref}}(t)=\bar{P}_{\textrm{ref}}(t)+\tilde{P}_{\textrm{ref}}(t),
\end{equation}
where
\begin{subequations}\label{eq:P_terms}
\begin{equation}\label{subeq:P_bar}
    \bar{P}_{\textrm{ref}}(t)=\sum_n
    \frac{\left|2\pi\Delta\omega\bar{V}_{\textrm{inc}}(\omega_n)\right|^2}
    {Z_0}\frac{\gamma_n^2}{\Delta\omega^2}e^{-2\gamma_n t},
\end{equation}
\begin{equation}\label{subeq:P_tilde}
    \tilde{P}_{\textrm{ref}}(t)=\sum_{n,m\neq
    n}\frac{\left|2\pi\Delta\omega\right|^2\bar{V}_{\textrm{inc}}(\omega_n)
    \bar{V}_{\textrm{inc}}^{*}(\omega_m)}{Z_0}\frac{\gamma_n\gamma_m}
    {\Delta\omega^2}e^{j(\omega_n-\omega_m^*) t}e^{2j(\psi_m-\psi_n)},
\end{equation}
\end{subequations}
and $\psi_n$ is the phase of $Z_0+j X_{n}$.

The two contributions to the reflected power (\ref{subeq:P_bar})
and (\ref{subeq:P_tilde}) are very different. In the first
contribution the terms decay exponentially and smoothly and the
sum is always positive. In fact, if we smooth over a timescale
longer than the Heisenberg time, this first term will remain
essentially unchanged.  The second term, on the other hand,
oscillates rapidly on a timescale comparable to the Heisenberg
time, but tends to zero if averaged over long timescales. For the
very long timescales needed to see the transition from
self-averaging to exponential decay, we can treat the rapidly
fluctuating terms in $P_{\textrm{ref}}(t)$ as random variables
with the phases in the exponents
($\left(\omega_n-\omega_m^{*}\right)t$) being uniformly
distributed. Under this assumption, we find that, for a single
realization of the chaotic cavity, the fluctuating part of
$P_{\textrm{ref}}$ is random and has a variance of
\begin{equation}\label{eq:P_ref_std_dev}
    \sigma^2=\langle\left[\tilde{P}_{\textrm{ref}}(t)\right]^2
    \rangle_{t}\leq \bar{P}^2_{\textrm{ref}}(t).
\end{equation}
where $\langle\ldots\rangle_{t}$ indicates a sliding averaging in
$t$ over a timescale that is long compared to the Heisenberg time
but short compared to the characteristic time for variation of
$\bar{P}_{\textrm{ref}}(t)$. That is, the order of magnitude of
the oscillating part of $P_{\textrm{ref}}$ is typically the same
as that of the smoothed part. Thus, if the smoothed part of
$P_{\textrm{ref}}$ drops exponentially, the fluctuations around it
will as well. Hence, if the power stays self-averaged, the
fluctuations will be as large as the signal itself.  When we
consider the transition from self-averaging to exponential decay,
we consider only the statistics of the smoothed part of
$P_{\textrm{ref}}$, ignoring the oscillating part which does not
contribute to the self-averaging. Thus in our theory we consider
only the time-averaged power $\bar{P}_{\textrm{ref}}(t)$, Eq.
(\ref{subeq:P_bar}), which is the key result of this section.

\section{Finding $P_{\gamma}(\gamma_n)$}
\label{sec:Finding_P_gamma}

From Eq.~\eqref{subeq:P_bar}, we see that the average reflected
power is a sum over contributions from exponentially decaying
modes. Because of the exponential decay, the relative amplitudes
of the modes will separate exponentially in time, with the modes
with the smallest $\gamma_n$ eventually dominating the sum.  Thus,
the crossover time from self-averaging to exponential decay
depends on the behavior of the probability distribution function
of $\gamma_n$ for small values of $\gamma_n$. In this section we
find the behavior of $P_{\gamma}(\gamma_n)$, the probability
distribution function for the decay rates for $\gamma_n\ll
\Delta\omega$. Previous work has been done on the subject (for
instance, in the case of a lasing chaotic cavity, see
Refs.~\cite{Schomerus_Petermann_factor,Schomerus_Quantum_linewidth}),
including analytical solutions for the $P_{\gamma}(\gamma)$ for
all $\gamma$
\cite{Fyodorov_Sommers_P_gamma,Fyodorov_Sommers_Titov_P_gamma},
but because we focus on the single port case with time reversal
symmetry for small $\gamma$ only, many approximations can be made
which greatly simplify the derivation, which we present here.

We start by considering the statistics of $\xi_n$, where $\xi_n$
is defined in Eq.~\eqref{eq:X_n_decomposed}. We show in Appendix
\ref{app:xi_dist} that the statistics of $\xi_n$ are given in
terms of the angle $\psi_n=\tan^{-1}(\xi_n)$, where $\psi_n$ is
distributed according to the pdf,
\begin{equation}\label{eq:tilde_phi_stats}
    P_{\psi_n}(\psi_n)=\frac{\cos(\psi_n)}{2}.
\end{equation}

Using this result and Eq.~\eqref{eq:gamma_n_solution}, we find an
expression for $P_{\gamma}(\gamma_n)$ where $\gamma_n\ll
\Delta\omega$:
\begin{equation}\label{eq:tilde_w_pdf}
    P_{\gamma}(\gamma_n)=\frac{1}{\sqrt{2\pi}}\int_{-\pi/2}^{\pi/2}
    d\psi_n\,\cos\psi_n\int_0^{\infty}
    dw\,e^{-w^2/2}\delta\left(\gamma_n-w^2\frac{r_{r}
    \Delta\omega}{\pi\left[1+(r_{r}\tan(\psi_n)+x_{r,n})^2\right]}\right),
\end{equation}
where $r_{r}=R_{Rad}(k)/Z_0$ and $x_{r}=X_{Rad}(k)/Z_0$. The
innermost integral can be evaluated leaving only an integral over
$\psi_n$.  Further, since we are only interested in the case of
small $\gamma_n\ll \Delta\omega$, the main contribution comes from
$|w|\ll 1$. The result is
\begin{equation}\label{eq:gamma_n_small_pdf}
    P_{\gamma}(\gamma_n)\cong
    \frac{P_0}{2\sqrt{\gamma_n\Delta\omega}}\textrm{ for }
    \gamma_n\ll \Delta\omega,
\end{equation}
where
\begin{equation}\label{eq:P_0_def}
    P_0=(2 r_{r})^{-1/2}\int_{-\pi/2}^{\pi/2}d\psi_n\,\sqrt{\cos^2
    \psi_n+(r_{r}\sin\psi_n+x_{r}\cos\psi_n)^2}.
\end{equation}

The quantity $P_0$ given in Eq.~\eqref{eq:P_0_def} can be
rewritten in terms of the radiation reflection coefficient of the
port that applies when the walls of the cavity have been moved out
to infinity,
\begin{equation}\label{eq:rho_r_defintion}
    \rho_r=\frac{z_r-1}{z_r+1},
\end{equation}
where $z_{r}=r_{r}+ix_{r}=(R_{Rad}+j X_{Rad})/Z_0$ is the
normalized radiation impedance of the antenna.  To see this, we
introduce the intermediate variable $\beta=z_{r}^2-1$ and define a
new integration variable $\phi=\psi_n-\textrm{arg}(\beta)/2$ in
Eq.~\eqref{eq:P_0_def}. The result of these variable changes is
\begin{equation}\label{eq:tilde_phi_w_n_equal_0_s3}
    P_0=\sqrt{2
    \frac{1-|\rho_{r}|}{1+|\rho_{r}|}}E\left(\frac{2j\sqrt{|\rho_{r}|}}
    {1-|\rho_{r}|}\right),
\end{equation}
where
\begin{equation}\label{eq:elliptic_2_def}
    E(k)=\int_{0}^{\pi/2}d\phi\,\sqrt{1-k^2\sin^2(\phi)}
\end{equation}
is the complete elliptic integral of the second kind.

We confirm Eqs.~(\ref{eq:gamma_n_small_pdf}) and
(\ref{eq:tilde_phi_w_n_equal_0_s3}) numerically by generating an
ensemble of $\gamma_n$ values. To do this we solve
Eq.~\eqref{eq:Z_sum_zero} by generating different realizations of
the Gaussian random variables $w_n$ and random matrix eigenvalues
$k_n^2$ appearing in the definition of $Z(\omega)$,
Eq.~\eqref{eq:statistical_impedance}.  We find the mode
frequencies by noting that as $Z_0\rightarrow\infty$,
$\omega_n\rightarrow k_n c$ for all modes.  We then introduce
$Y_0=Z_0^{-1}$ and differentiate both sides of
Eq.~\eqref{eq:Z_sum_zero} with respect to $Y_0$, obtaining a
differential equation for $\omega_n(Y_0)$,
\begin{equation}\label{eq:w_n_diffeq}
    \frac{d\omega_n}{d Y_0}=\frac{Z^2(\omega_n)}{Z'(\omega_n)},
\end{equation}
which can be solved numerically to find $\omega_n$ for finite
$Z_0$. Note that although both $Z^2(\omega_n)$ and $Z'(\omega_n)$
are singular as $\omega_n\rightarrow k_n c$, their ratio is
finite.

\begin{figure}
\includegraphics{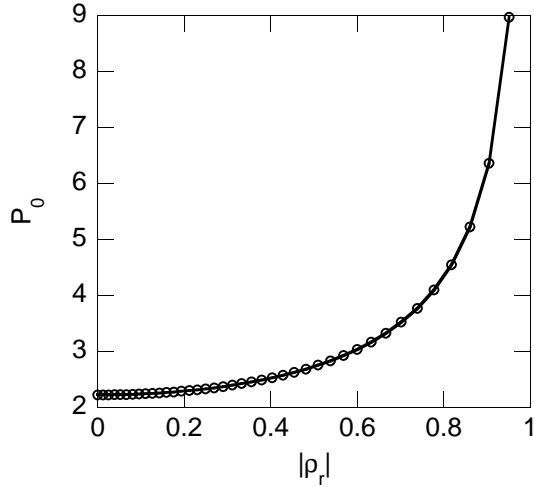}
\caption{\label{fig:P_wn_0} A comparison of numerically generated
values for $P_0$ (circles) with the theoretical result from
Eq.~\eqref{eq:tilde_phi_w_n_equal_0_s3} (the solid line). The
circles represent numerical calculations of $P_0$ with the
radiation reactance of the port set to be $X_{Rad}=0$.  To get
different values of $|\rho_r|$, $Y_0$ was changed as described in
Eq.~\eqref{eq:w_n_diffeq}.}
\end{figure}

By generating 1000 different realizations of $k_n^2$ and
$\omega_n$ (truncating the spectrum to include only 600 terms),
and integrating Eq.~\eqref{eq:w_n_diffeq} numerically using
fourth-order Runga-Kutta from $Y_0=0$ to $Y_0=R_{Rad}^{-1}$, it is
possible to generate pdfs of $\tilde{w}_n\equiv\sqrt{\gamma_n}$ as
a function of $|\rho_r|$. We choose the pdfs of $\tilde{w}_n$
instead of $\gamma_n$ because
$P_{\tilde{w}}(\tilde{w}=0)=P_0/\Delta\omega$, which is finite and
thus numerically easier to fit.  The results are shown in
Fig.~\ref{fig:P_wn_0} where the numerical results and the theory
are seen to be in clear agreement.  We note that this numerical
test (solving Eq.~\eqref{eq:w_n_diffeq} for $Y_0=R_{Rad}^{-1}$)
does not assume the weak coupling limit and thus confirms our
assumptions in obtaining Eq.~\eqref{eq:tilde_phi_w_n_equal_0_s3}.

\section{The Statistics of $\bar{P}_{\textrm{ref}}(t)$}
\label{sec:bar_P_stats}

The smoothed reflected power $\bar{P}_{\textrm{ref}}(t)$ given by
Eq.~\eqref{subeq:P_bar} is a sum of terms each of which is a
random variable. The terms are not strictly independent.  This
follow from the fact that there are correlations between the
eigenvalues of the closed system, and $\gamma_n$, given by
Eq.~\eqref{eq:gamma_n_solution}, depends on these eigenvalues
through the reactance $X_n$, defined in Eq.~\eqref{eq:X_n_def}.
Fortunately the correlation is significant only for almost
adjacent modes. For times large enough that the self-averaging
breaks, the fraction of modes contributing will be small, and
thus, the majority of contributing modes will be well separated
and approximately independent of each other.

Hence for our purposes, $\bar{P}_{\textrm{ref}}$ can be treated as
a sum of a large number of independent terms. Thus, for times when
a large number (but small fraction) of modes have comparable
magnitudes, for an ensemble of cavity realizations,
$\bar{P}_{\textrm{ref}}$ is a Gaussian random variable centered on
$\langle \bar{P}_{\textrm{ref}}(t)\rangle$ with a small standard
deviation. As we demonstrate in the following sections, the
standard deviation starts out small, but as the number of
contributing modes decreases, the standard deviation increases
relative to the mean, eventually becoming much larger than the
mean.  As this happens, the simple Gaussian distribution changes
into a more complex distribution with the majority of modes
becoming much smaller than the average, corresponding to the shift
from self-averaging to exponential decay.

These shifts can be treated analytically by considering the
moments of $\bar{P}_{\textrm{ref}}$. We first (Sec.~
\ref{subsec:mean_and_variance}) consider the mean and standard
deviation of $\bar{P}_{\textrm{ref}}$ to find a scaling law
describing the transition from Gaussian to non-Gaussian behavior.
Armed with the results from this comparison, in
Sec.~\ref{subsec:higher_moments} we generalize the results to
higher-order moments (via the cumulants), showing that for large
times all moments of $\bar{P}_{\textrm{ref}}$ obey the same
scaling law. We then numerically demonstrate that the cumulative
distribution function of $\bar{P}_{\textrm{ref}}/\langle
\bar{P}_{\textrm{ref}}\rangle$ satisfies the scaling law for
multiple pulse shapes, as predicted.

\subsection{The Mean and Variance}
\label{subsec:mean_and_variance} We can calculate the mean and the
variance of $\bar{P}_{\textrm{ref}}$ for all times as
\begin{equation}\label{eq:P_ref_avg}
    \langle \bar{P}_{\textrm{ref}}\rangle=\sum_{n} \frac{|2\pi \Delta\omega
    \bar{V}_{\textrm{inc}}(\omega_n)|^2}{Z_0}\mu_1,
\end{equation}
and
\begin{equation}\label{eq:P_ref_sigma}
    \langle (\bar{P}_{\textrm{ref}}-\langle \bar{P}_{\textrm{ref}}\rangle)^2
    \rangle=\sum_{n} \frac{|2\pi \Delta\omega
    \bar{V}_{\textrm{inc}}(\omega_n)|^4}{Z_0^2}(\mu_2-\mu_1^2),
\end{equation}
where
\begin{equation}\label{eq:mu_def}
    \mu_m(t)=\int_0^{\infty} \frac{d\gamma
    P_0}{2\sqrt{\gamma\Delta\omega}}\left[\frac{\gamma^2}{\Delta\omega^2}
    e^{-\gamma t}\right]^m.
\end{equation}
Evaluation of the integral in Eq.~\eqref{eq:mu_def} gives
\begin{equation}\label{eq:mu_full_form}
    \mu_m(t)=\frac{P_0}{2 (m \Delta\omega
    t)^{2m+1/2}}\Gamma(2m+1/2).
\end{equation}
Equations~\eqref{eq:P_ref_avg} and \eqref{eq:mu_def} give the
result that the average reflected power (averaged over an ensemble
of reflecting cavities) decreases as a power law in time, which is
in agreement with previous theory
\cite{Dittes_Harney_Muller,Dittes_Long_Decay},
\begin{equation}\label{eq:P_ref_power_law_explicit}
\langle\bar{P}_{\textrm{ref}}(t)\rangle\sim t^{-5/2}.
\end{equation}

Equation~\eqref{eq:P_ref_sigma} is useful for finding the range of
values that are most likely to contain $\bar{P}_{\textrm{ref}}$;
for small times with an approximately Gaussian pdf for
$\bar{P}_{\textrm{ref}}$, we expect that the majority of
realizations will fall within the range $[\langle
\bar{P}_{\textrm{ref}}\rangle-2\sigma_{P},\langle
\bar{P}_{\textrm{ref}}\rangle+2\sigma_{P}]$ where
$\sigma_P=\langle (\bar{P}_{\textrm{ref}}-\langle
\bar{P}_{\textrm{ref}}\rangle)^2\rangle^{1/2}$.  For large times,
however, $\sigma_P>\langle \bar{P}_{\textrm{ref}}\rangle$.  We see
this by first considering the ratio
\begin{equation}\label{eq:mean_std_dev_ratio}
    \frac{\mu_2}{\mu_1^2}=\frac{(\Delta\omega
    t)^{1/2}}{P_0}\frac{\Gamma(9/2)}{2^{7/2}\Gamma(5/2)^2}.
\end{equation}
Thus, for large times, $\mu_2\gg \mu_1^2$, and $\mu_2$ dominates
Eq.~\eqref{eq:P_ref_sigma}.  For large times, we have
\begin{equation}\label{eq:sigma_P_avg_rat}
    \frac{\sigma_P^2}{\langle
    \bar{P}_{\textrm{ref}}\rangle^2}=\frac{(t\Delta\omega)^{1/2}}
    {P_0}\frac{\Gamma(9/2)}{2^{7/2}\Gamma(5/2)^2}\frac{\sum_n |V_{\textrm{inc}}
    (\omega_n)|^4}{\left[\sum_n |V_{\textrm{inc}}(\omega_n)|^2\right]^2}
\end{equation}
Equation~\eqref{eq:sigma_P_avg_rat} can be made more transparent
by considering the sums over $|V_{\textrm{inc}}|^{2m}$.  The
incident pulse can be considered to have two independent
properties: a shape and a width.  If we double the width of the
pulse in the frequency domain (or equivalently if we halve the
average mode separation) without changing the shape, the sums in
Eq.~\eqref{eq:sigma_P_avg_rat} will, to a good approximation,
simply double.  We thus define the effective number of modes
excited by the wave to be
\begin{equation}\label{eq:N_def}
    N=\frac{\left[\sum_n |V_{\textrm{inc}}(\omega_n)|^2\right]^2}{\sum_n
    |V_{\textrm{inc}}(\omega_n)|^4}.
\end{equation}
In the case of a square wave excitation in the frequency domain,
Eq.~\eqref{eq:N_def} gives exactly the number of modes excited. In
the case of more typical pulses, such as a Gaussian pulse,
Eq.~\eqref{eq:N_def} defines a relationship between the pulse
width and the number of significant excited modes.

Substituting Eq.~\eqref{eq:N_def} into
Eq.~\eqref{eq:sigma_P_avg_rat}, we get
\begin{equation}\label{eq:sigma_P_avg_rat_normed}
    \frac{\sigma_P^2}{\langle
    \bar{P}_{\textrm{ref}}\rangle^2}=\tau^{1/2}\frac{\Gamma(9/2)}
    {2^{7/2}\Gamma(5/2)^2},
\end{equation}
where
\begin{equation}\label{eq:tau_def}
    \tau=\frac{t \Delta\omega}{N^2P_0^2}.
\end{equation}
As long as $\sigma_P/\langle \bar{P}_{\textrm{ref}}\rangle$ is
small, it is reasonable to expect the majority of realizations of
$\bar{P}_{\textrm{ref}}$ to be within two sigma of the average,
and numerically we find that this is true. From
Eq.~\eqref{eq:sigma_P_avg_rat_normed}, we see that for
$t\Delta\omega\gg 1$ and $\tau\ll 1$ (possible because $N$ is
assumed to be large) this is possible.  Eventually the standard
deviation will be comparable to the mean and for very long times
the standard deviation will be much larger than the mean.  This
shift corresponds to the change from self-averaging to exponential
decay.

\subsection{Higher Moments}
\label{subsec:higher_moments}

An analysis of the higher moments of $\bar{P}_{\textrm{ref}}$
follows essentially the same steps as those to find the mean and
variance. We find the moments of $\bar{P}_{\textrm{ref}}$ by
finding the moments of the individual terms in
$\bar{P}_{\textrm{ref}}$, dropping all but the leading order term
in $t^{-1/2}$, and combining them properly to get the moments of
the sum.  We cannot do this by simply summing the moments of the
individual terms; the sums of the moments are not in general the
moments of the sum. However, if we define the moment-generating
function,
\begin{equation}\label{eq:moment_function}
    M(h)=\langle e^{h
    \bar{P}_{\textrm{ref}}}\rangle=1+\sum_{p=1}^{\infty}\frac{h^p\langle
    \bar{P}_{\textrm{ref}}^p\rangle}{p!},
\end{equation}
we see that the moments of $\bar{P}_{\textrm{ref}}$ are given by
\begin{equation}\label{eq:moment_from_generator}
    \langle \bar{P}_{\textrm{ref}}^m\rangle=M^{(m)}(0).
\end{equation}
Here $M^{(m)}(h)$ is the $m$th derivative of $M(h)$ with respect
to its argument.  This can be related to a function known as the
cumulant-generating function
\begin{equation}\label{eq:cumulant_function}
    g(h)=\log(M(h))=\sum_{p=1}^{\infty}\kappa_p \frac{h^p}{p!}
\end{equation}
where $\kappa_m$ is the $m$th cumulant, defined as
\begin{equation}\label{eq:sum_cumulant_def}
    \kappa_m=g^{(m)}(0).
\end{equation}

We show in Appendix (\ref{app:cumulant_appendix}) that, in analogy
to Eq.~\eqref{eq:sigma_P_avg_rat}, the higher-order cumulants (and
thus all higher-order moments) of $\bar{P}_{\textrm{ref}}$ are
given by
\begin{equation}\label{eq:normed_kappa}
    \frac{\kappa_m}{\kappa_1^m}=\left(2
    \sqrt{\tau}\right)^{m-1}\frac{\Gamma(2m+1/2)}{m^{2m+1/2}\Gamma(5/2)^m}
    \frac{N^{m-1}\sum_n|V_{\textrm{inc}}(\omega_n)|^{2m}}{\left(
    \sum_n|V_{\textrm{inc}}(\omega_n)|^{2}\right)^m}.
\end{equation}
If we use the definition of $N$ from Eq.~\eqref{eq:N_def} and
approximate all sums over $n$ with integrals over $\omega_n$, we
find that the expression
$N^{m-1}\sum_n|V_{\textrm{inc}}(\omega_n)|^{2m}/
\left(\sum_n|V_{\textrm{inc}}(\omega_n)|^{2}\right)^m$
is, to a good approximation, independent of the width of the power
spectrum but dependent on the shape.  In the case of a square
power spectrum, this factor is identically one for all $m$.  For a
Gaussian pulse we find that
\begin{equation}\label{eq:Gauss_N_ratio}
    \frac{N^{m-1}\sum_n|V_{\textrm{inc}}(\omega_n)|^{2m}}{\left(
    \sum_n|V_{\textrm{inc}}(\omega_n)|^{2}\right)^m}=\sqrt{\frac{2^{m-1}}{m}}.
\end{equation}
Similarly, for a pulse with a Lorentzian power spectrum,
\begin{equation}\label{eq:Lorentz_N_ratio}
    \frac{N^{m-1}\sum_n|V_{\textrm{inc}}(\omega_n)|^{2m}}{\left(
    \sum_n|V_{\textrm{inc}}(\omega_n)|^{2}\right)^m}=\frac{2^{m-1}
    \Gamma(m-\frac{1}{2})}{\sqrt{\pi}\Gamma(m)}.
\end{equation}

Equation~\eqref{eq:normed_kappa}, combined with replacing the sums
over $|V_{\textrm{inc}}(\omega)|^{2m}$ with integrals,
demonstrates the most important theoretical result of this paper:
\textit{all statistical properties of the reflected power depend
only on the shape of the pulse (independent of width) and the
normalized time $\tau$ defined in Eq.~\eqref{eq:tau_def}}. Thus
the cross-over from self-averaging to exponential decay, no matter
how measured, will depend only on $\tau$ and the pulse shape.

\section{Numerical Results}\label{sec:numerics}

In this section, we compare different methods of calculating
$\bar{P}_{\textrm{ref}}(t)$ to show that our theoretical
conclusions are correct. To view the resulting distributions, we
find the ensemble average of the calculated values of
$\bar{P}_{\textrm{ref}}(t)$ and then compare the individual
realizations to the average.  In particular, we define
$C(\alpha,\tau)$ to be the fraction of realizations which are less
than $\alpha$ times the ensemble average (i.e. $C(\alpha,\tau)$ is
the cumulative distribution of $\bar{P}_{\textrm{ref}}$ at the
normalized time $\tau$).

\begin{figure*}
\includegraphics{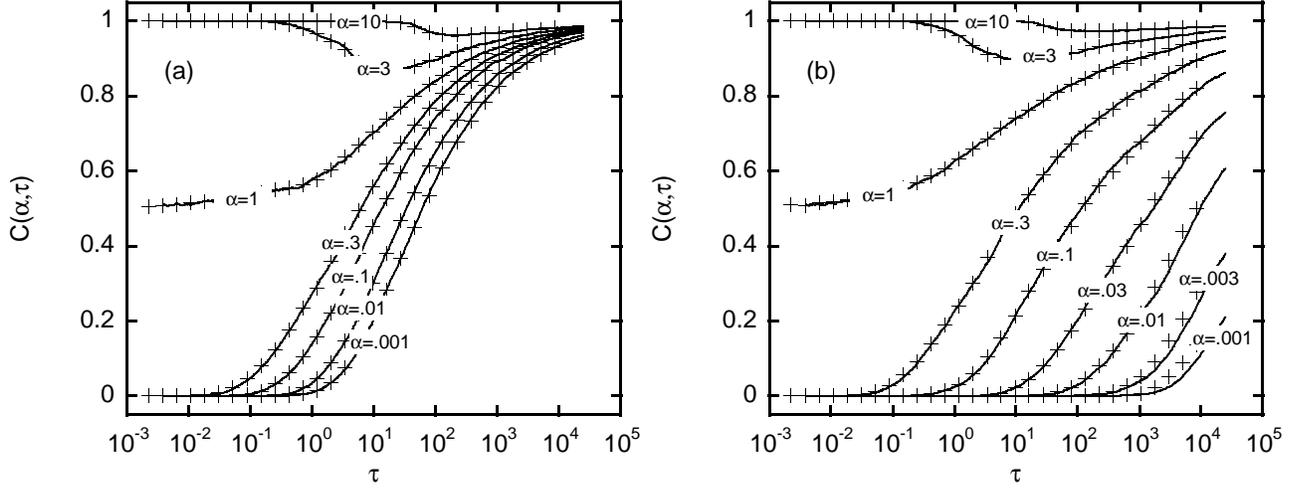}
\caption{\label{fig:CDF_GaussLorentz} The fraction of realizations
of $\bar{P}_{\textrm{ref}}$ which are less than $\alpha\langle
\bar{P}_{\textrm{ref}}\rangle$ as a function of normalized time
$\tau$ for (a) a Gaussian spectrum and (b) a Lorentzian spectrum.
The black lines(`+' symbols) represent the statistics for
$N=20(30)$. Note that plots for $N=20$ and $N=30$ are slightly
different for the Lorentzian case with small $\alpha$. This is due
to the fact that the contributions for small $\alpha$ come from
the tails of the distribution, which we numerically truncated to
calculate these plots.}
\end{figure*}

To both test and evaluate the theoretical results in
Sec.~\ref{sec:bar_P_stats}, we perform two separate, independent
calculations which should, according to our theory, produce the
same results.  The first method calculates the sum in
Eq.~\eqref{subeq:P_bar} with the $\gamma_n$ independent of
$\textrm{Re}(\omega_n)$ and distributed according to the
Porter-Thomas distribution with one degree of freedom,
\begin{equation}\label{eq:test_gamma}
    P(\gamma)=\frac{e^{-\gamma/2}}{\sqrt{2\pi\gamma}}.
\end{equation}
This distribution is chosen because it has the same behavior for
small $\gamma$ as is indicated in
Eq.~\eqref{eq:gamma_n_small_pdf}.  We consider two different pulse
spectra, $\bar{V}_{\textrm{inc}}(\omega_n)$, Gaussian and
Lorentzian, with two different widths $N=20$ and $30$, where $N$
is defined in Eq.~\eqref{eq:N_def}.  Finally, we take the
$\omega_n$ to be uniformly spaced when evaluating the sums.  We
call these results the theoretical results because they are a
numerical evaluation of the theoretical assumptions used in
Sec.~\ref{sec:bar_P_stats}. The theoretical results are shown in
Fig.~\ref{fig:CDF_GaussLorentz} for the case of the two pulse
shapes and two spectral widths.  The first thing to note about the
plots is that the results for $N=20$ and $N=30$ lie on top of each
other, showing that the definition of $\tau$(\ref{eq:tau_def})
correctly accounts for variation of the pulse width.  (There is a
small deviation in the Lorentzian case for small values of
$\alpha$ that will be addressed subsequently.)  The second thing
to note is that the $C(\alpha>0.3)$ curves for the two pulse
shapes are very similar.  Thus, the fraction of realizations close
to or greater than the mean is the same in the two cases.  Where
the two pulse shapes differ is for times $\tau\gg 1$ and small
$\alpha\ll 1$.  In the Gaussian case almost all realizations fall
well below the average as $\tau$ gets large, whereas in the
Lorentzian case there is a larger fraction of realizations with
measurable power ($\alpha>0.001$) at late time.  This is due to
the long tail in the Lorentzian distribution exciting a large
number of modes with small but significant levels of power.  The
difference between the $N=20$ and $N=30$ cases is due to
truncation of the spectrum at $600$ modes.

The second test employs the time-domain code used to generate the
data in Fig.~\ref{fig:single_realizations}.  We then time-smooth
the resulting power (using a Gaussian window with a width of $10$
Heisenberg times) to calculate $\bar{P}_{\textrm{ref}}$. The time
domain code is described in Appendix~\ref{app:time_domain_code}.
In Fig.~(\ref{fig:time_domain_vs_theory}) we compare results for
$C(\alpha,\tau)$ using 50 realizations with the theoretical
curves. The time-domain code is run only to $\tau=1$ which for
these parameters corresponds to $~1744$ Heisenberg times.  The
time domain simulation results agree well with the theoretical
results considering the finite sample size.
\begin{figure}
\includegraphics{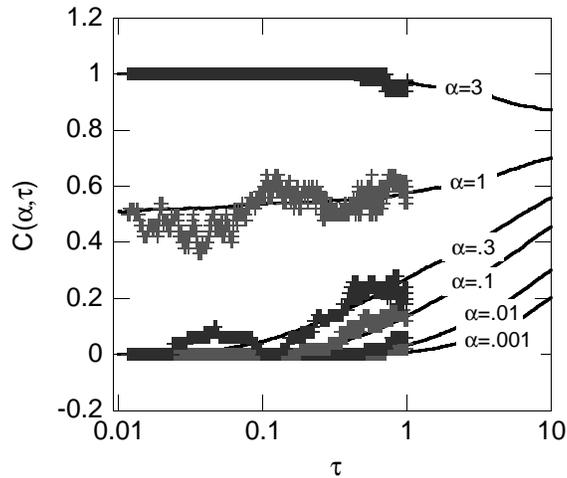}
  \caption{The fraction of realizations
of $\bar{P}_{\textrm{ref}}$ which are less than $\alpha\langle
\bar{P}_{\textrm{ref}}\rangle$ as a function of normalized time
$\tau$ for the theoretical results calculated numerically (the
solid lines) and the same results calculated from integrating
Eq.~\eqref{eq:startEq} directly (indicated by the `+' symbols).
Random Matrix Theory is explicitly used to calculate the spectrum
and coupling constants for the time-domain
integration.}\label{fig:time_domain_vs_theory}
\end{figure}

In addition, we have performed tests which have allowed the value
of $P_0$ to vary, and have solved Eq.~\eqref{eq:w_n_diffeq} to get
the complex values of $\omega_n$. The results agree well with the
theoretical results of Fig.~\ref{fig:CDF_GaussLorentz} and are not
displayed.

\section{Conclusions}
\label{sec:conclusions} In this paper, we have found numerically
and theoretically that the long term behavior of power reflected
from a lossless, microwave cavity excited through a single port
self-averages for times larger than the Heisenberg time, decaying
as a power law in time.  We have also found, theoretically and
numerically, that for times much longer than the Heisenberg time,
when $\tau$, the normalized time, is of order 1, that single modes
in the cavity will begin to dominate the long term decay and the
reflected power will begin to decay exponentially.  The details of
this behavior have been found to depend on the shape of the power
spectrum of the incident pulse that excited the cavity, but to
otherwise depend only on the normalized time. Because much of the
theory used to derive this behavior depends only on generic Random
Matrix Theory, we expect that this behavior will translate into
other lossless wave-chaotic systems (e.g., acoustic, quantum
mechanical, etc.), independent of details.

\acknowledgements

We would like to thank Dr.\ S.\ M.\ Anlage and Dr.\ R.\ E.\ Prange
for helpful discussions. This work was supported by the
USAFOSR~grant~\#FA95500710049.

\appendix
\section{Finding the Distribution of $\xi_n$ for Small $\gamma_n$}
\label{app:xi_dist}

To find the distribution of $\xi_n$ defined in
Eq.~\eqref{eq:X_n_decomposed} for small $\gamma_n$, we exploit the
fact that, in a two-port system with the ports identical and
described by Random Matrix Theory, the diagonal elements of the
normalized impedance matrix each have the same statistics as the
single-port normalized impedance. Then using the exact statistics
of the two-port RMT impedance, we can find the statistics of the
one-port impedance (\ref{eq:X_n_def}).

We see this by first writing the elements of the two-port
normalized impedance matrix as a sum, analogous to
Eq.~\eqref{eq:statistical_impedance},
\begin{equation}\label{eq:two_port_xi_sum}
    \xi_{i,j}=-\frac{j}{\pi}\sum_n\frac{w_{i,n}w_{j,n}}{k^2-k_n^2},
\end{equation}
where the $w_{i,n}$ are independent Gaussian random variables and
the $k_n^2$ have the statistics of the eigenvalues of a GOE random
matrix.

As shown in previous work \cite{Henry_paper_many_port}, the 2x2
matrix $\xi$ has the following statistics: its eigenvalues
$\tan\theta_1$, and $\tan\theta_2$ have a joint pdf,
\begin{equation}\label{eq:joint_theta_pdf}
    P(\theta_1,\theta_2)\propto
    \left|\sin\left(\frac{\theta_2-\theta_1}{2}\right)\right|,
\end{equation}
and its eigenvectors $(\cos\nu,\sin\nu)$ and $(-\sin\nu,\cos\nu)$
have $\nu$ uniformly distributed and independent of $\theta_1$ and
$\theta_2$. Consequently, a diagonal element of $\xi$ can also be
parameterized as
\begin{equation}\label{eq:two_port_xi_direct}
    \xi_{i,i}=\cos^2\eta\tan\theta_1+sin^2\eta\tan\theta_2.
\end{equation}

Comparing Eqs. (\ref{eq:two_port_xi_sum}) and
(\ref{eq:two_port_xi_direct}), we see that the singularity at
$k=k_n$ in Eq.~\eqref{eq:two_port_xi_sum} is matched by either
$\theta_1$ or $\theta_2$ going through $\pi/2$; for specificity we
assume that it is $\theta_1$. For small $\gamma_n$, corresponding
to small $w_n^2$, the coefficient of the singularity is small,
which corresponds to $\cos^2\eta\approx 0$. Thus, for small
$\gamma_n$, $\xi_n$ has the statistics given by
\begin{equation}\label{eq:xi_cos_equality}
    \xi_n=\tan\theta_2|_{\theta_1=\pi/2}
\end{equation}
which inserted into Eq.~\eqref{eq:joint_theta_pdf} produces the
pdf for $\psi_n=\tan^{-1}\xi_n=\theta_2$
\begin{equation}\label{eq:xi_pdf_found}
    P(\psi_n)=\frac{\cos\psi_n}{2}
\end{equation}

\begin{figure}
\includegraphics{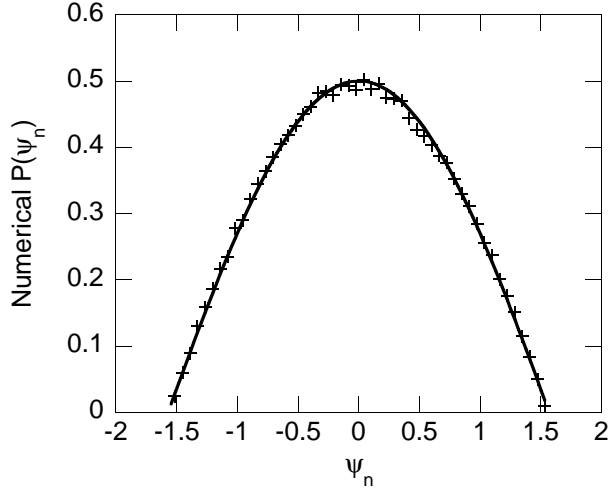}
\caption{\label{fig:pdf_psi} A comparison of a
numerically-generated pdf of $\psi_n$ (`+' symbols) with the
anticipated result from Eq.~\eqref{eq:xi_pdf_found},
$\cos(\psi_n)/2$ (the solid line).}
\end{figure}

Numerically we confirm this by generating a single 600x600 element
matrix from the Gaussian Orthogonal Ensemble and calculating and
scaling the eigenvalues to get an appropriate spectrum. We then
repeatedly generate 600 realizations of 600 coupling constants and
use them to calculate 360,000 realizations of $X_n$,  which we
then normalize to calculate $\psi_n$.  The resulting statistics
are demonstrated in Fig.~\ref{fig:pdf_psi}.

\section{Finding the Cumulants of $\bar{P}_{\textrm{ref}}$}

\label{app:cumulant_appendix}

To obtain Eq.~\eqref{eq:normed_kappa}, we note that the cumulant
generating function of $\bar{P}_{\textrm{ref}}$ obeys
\begin{eqnarray}
    g(h)&=&\log\left(\left\langle \exp\left(h
    \sum_n
    \frac{\left|2\pi\Delta\omega\bar{V}_{\textrm{inc}}(\omega_n)\right|^2}
    {Z_0}\frac{\gamma_n^2}{\Delta\omega^2}e^{-2\gamma_n
    t}\right)\right\rangle\right)\nonumber\\\label{eq:cumulative_property}
    &=&\sum_n\log\left(\left\langle \exp\left(h
    \frac{\left|2\pi\Delta\omega\bar{V}_{\textrm{inc}}(\omega_n)\right|^2}
    {Z_0}\frac{\gamma_n^2}{\Delta\omega^2}e^{-2\gamma_n
    t}\right)\right\rangle\right).
\end{eqnarray}
This result is a specific example of a general property of
cumulants \cite{cumulant_book}: The $m$th cumulant of a sum of
independent variables is the sum of the $m$th cumulants of the
single variables. Thus, in analogy to
Eq.~\eqref{eq:cumulant_function}, we define the
cumulant-generating function and the cumulants $\tilde{\kappa}_p$
for each term in the sum in Eq.~\eqref{eq:cumulative_property} as
\begin{equation}\label{eq:cum_for_terms}
    \tilde{g}(q)=\log\left(\left\langle \exp\left(q\frac{\gamma_n^2}
    {\Delta\omega^2}e^{-2\gamma_n
    t}\right)\right\rangle\right)=\sum_{p=1}^{\infty}
    \tilde{\kappa}_{p}\frac{q^p}{p!},
\end{equation}
and by matching coefficients of $h^m$ in
Eq.~\eqref{eq:cumulative_property}, we get that
\begin{equation}\label{eq:kappa_as_sum}
    \kappa_m=\tilde{\kappa}_m\sum_n\frac{\left|2\pi\Delta\omega
    \bar{V}_{\textrm{inc}}(\omega_n)\right|^{2m}}{Z_0^m}.
\end{equation}
All that is left is to find the long-term behavior for
$\tilde{\kappa}_m$.  To do this, we note that we can rewrite the
average exponential in Eq.~\eqref{eq:cum_for_terms} as
\begin{equation}\label{eq:exponential_rewrite}
    \left\langle \exp\left(q\frac{\gamma_n^2}{\Delta\omega^2}
    e^{-2\gamma_n t}\right)\right\rangle=
    1+\sum_{n=1}^{\infty}\frac{q^n\mu_n}{n!}.
\end{equation}
Because $q$ is a dummy variable which can be arbitrarily small, we
can also expand the logarithm in Eq.~\eqref{eq:cum_for_terms} to
get that
\begin{equation}\label{eq:matching_coefficients}
    \sum_{p=1}^{\infty}\tilde{\kappa}_{p}\frac{q^p}{p!}=
    \sum_{n=1}^{\infty}\frac{q^n\mu_n}{n!}
    -\frac{1}{2}\left(\sum_{n=1}^{\infty}\frac{q^n\mu_n}{n!}\right)^2
    +\frac{1}{3}\left(\sum_{n=1}^{\infty}\frac{q^n\mu_n}{n!}\right)^3-\ldots.
\end{equation}
By matching coefficients of $q^p$ on both sides of
Eq.~\eqref{eq:matching_coefficients}, we find that
\cite{cumulant_book},
\begin{eqnarray}\label{eq:term_cum_and_moment}
    \tilde{\kappa}_1&=&\mu_1,\\
    \tilde{\kappa}_2&=&\mu_2-\mu_1^2,\\
    \tilde{\kappa}_m&=&\mu_m-m
    \mu_{m-1}\mu_1+\ldots-(-1)^m\mu_1^m,\label{eq:kappa_from_mu_sum}
\end{eqnarray}
where the elided terms are products of different $\mu_n$ such that
the indices add up to $m$. For large $t$, all of these polynomial
terms are small compared $\mu_m$.  We can see this by noting that
$\mu_m\propto (t\Delta\omega)^{-2m-1/2}$. Thus
\begin{equation}\label{eq:kappa_ratio}
    \frac{\mu_m}{\mu_{l}\mu_{m-l}}\propto (t\Delta\omega)^{1/2},
\end{equation}
where the proportionality constant can be shown to be order 1. For
every extra factor of $\mu_l$ included in a term, we pick up an
extra factor of $(t\Delta\omega)^{1/2}$ in the numerator of the
ratio between $\mu_m$ and that term. Thus for large times we have
that $\mu_m$ is much greater than any of the other polynomial
terms in Eq.~\eqref{eq:kappa_from_mu_sum} and therefore
\begin{equation}\label{eq:kappa_mu_equivalence}
    \tilde{\kappa}_m\approx \mu_m.
\end{equation}
Combining Eqs.~(\ref{eq:mu_full_form}), (\ref{eq:kappa_as_sum}),
and (\ref{eq:kappa_mu_equivalence}), we get
Eq.~\eqref{eq:normed_kappa}.

\section{The time domain code}\label{app:time_domain_code}

In this section, we describe the time domain code used to create
the realizations in Fig.~\ref{fig:single_realizations}. This code
effectively solves Eq.~\eqref{eq:startEq} using the approximations
that were inserted into Eq.~\eqref{eq:impedance_as_sum} to produce
Eq.~\eqref{eq:statistical_impedance}.  In addition, it makes use
of a slowly varying envelope approximation which greatly increases
the size of the numerically stable time-step and also transforms
Maxwell's Equations into Schr\"{o}dinger's Equation.

To solve Eq.~\eqref{eq:startEq}, we first find expand $V(x,y,t)$
in terms of the eigenfunctions of the closed system
\begin{equation}\label{eq:v_t_sum}
    \bar{V}_{T}(x,y,t)=\sum_n \frac{\tilde{c}_{n}(t)\phi_{n}(x,y)}
    {\sqrt{\int d\theta\,|u(\vec{w}_0 c)|^2}}.
\end{equation}
We note that the $c_{n}(\omega)$ from Eq.~\eqref{eq:v_omega_sum}
are proportional to Fourier transforms of the $\tilde{c}_{n}(t)$.
Substituting Eq.~\eqref{eq:v_t_sum} into Eq.~\eqref{eq:startEq}
and using the orthonormality of the $\phi_n$, we get
\begin{equation}\label{eq:individual_terms_t_equation}
    \frac{1}{c^2}\frac{d^2}{dt^2}\tilde{c}_{n}(t)+k_n^2\tilde{c}_{n}(t)=
    \frac{8\pi R_{R}(\omega_0 c)}{\omega_0}\frac{d
    I(t)}{dt}\left[\frac{\int dx\,dy\,u\phi_n}
    {\sqrt{\int d\theta\,|u(\vec{w}_0 c)|^2}}\right],
\end{equation}
where we have used the definition of radiation resistance from
Ref.~\cite[Eq.~19]{Henry_paper_one_port} to remove the factor
$h\mu$. The value of $\omega_0$ is the modulation frequency used
in the envelope approximation (See
Eqs.~\eqref{eq:envelope_statement} and
\eqref{eq:extreme_inequality}).

To apply the envelope approximation, we assume that
\begin{subequations}\label{eq:envelope_statement}
\begin{equation}\label{eq:envelope_statement_a}
    I(t)=I_{\textrm{env}}(t)e^{j\omega_0 t}
\end{equation}
\begin{equation}\label{eq:envelope_statement_b}
    \tilde{c}_{n}(t)=d_{n}(t)e^{j\omega_0 t}
\end{equation}
\end{subequations}
where
\begin{subequations}\label{eq:extreme_inequality}
\begin{equation}\label{eq:extreme_inequality_a}
    \frac{d}{dt}I_{\textrm{env}}(t)\ll  \omega_0 I_{\textrm{env}}(t)\\
\end{equation}
\begin{equation}\label{eq:extreme_inequality_b}
    \frac{d}{dt}d_{m}(t) \ll  \omega_0 d_{m}(t).
\end{equation}
\begin{equation}\label{eq:extreme_inequality_c}
    \frac{d^2}{dt^2}d_{m}(t) \ll  \omega_0 \frac{d}{dt}d_{m}(t).
\end{equation}
\end{subequations}
Then we drop all terms which are small, noting that $k_n\approx
\omega_0/c$, which implies that $k_n^2 c^2-\omega_0^2=(k_n
c-\omega_0)(k_n c+\omega_0)$ is on the order of $\omega_0$. This
gives us
\begin{equation}\label{eq:individual_terms_envelope_equation}
    \left[\frac{2j\omega_0}{c^2}\frac{\partial}{\partial t}
    +(k_n^2-\frac{\omega_0^2}{c^2})\right]d_{n}(t)=
    8j\pi R_{R}(\omega_0 c)I_{\textrm{env}}(t)\left[\frac{\int dx\,dy\,u\phi_n}
    {\sqrt{\int d\theta\,|u(\vec{w}_0 c)|^2}}\right]
\end{equation}

Again we replace the overlap integral between $\phi_n$ and $u$
with the statistical approximation found in
Ref.~\cite[Eq.~14]{Henry_paper_one_port} to get
\begin{equation}\label{eq:individual_terms_t_equation_random}
    \left[\frac{2j\omega_0}{c^2}\frac{\partial}{\partial t}+
    (k_n^2-\frac{\omega_0^2}{c^2})\right]d_{n}(t)=
    \sqrt{8\Delta}w_nj R_{R}(\omega_0
    c)I_{\textrm{env}}(t).
\end{equation}

Similarly, combining Eqs.~\eqref{eq:v_t_sum} and \eqref{eq:V_def}
and using the envelope approximation throughout, we get
\begin{equation}\label{eq:V_time_def}
    V_{\textrm{env}}(t)=\sum_n V_n(t),
\end{equation}
where $V_{\textrm{env}}(t)$ is the envelope of $V(t)$ in analogy
to Eq.~\eqref{eq:envelope_statement} and
\begin{equation}\label{eq:V_n_def}
    V_n(t)=\frac{\sqrt{\Delta}}{4\pi} d_{n}(t)w_n.
\end{equation}

Solving Eqs.~(\ref{eq:V_sum}) and (\ref{eq:I_V_diff}) for $I(t)$
by eliminating $V_{\textrm{ref}}(t)$ and inserting the result into
Eq.~\eqref{eq:individual_terms_t_equation_random}, we get
\begin{equation}\label{eq:solved_I_time_equation}
    \left[\frac{2j\omega_0}{c^2}\frac{\partial}{\partial t}
    +(k_n^2-\frac{\omega_0^2}{c^2})\right]V_{n}(t)
    =j \frac{\Delta R_{R}(\omega_0 c) w_n^2}{\sqrt{2}\pi Z_0}
    \left(2 V_{i,\textrm{env}}(t)-\sum_m V_{m}(t)\right),
\end{equation}
where $V_{i,\textrm{env}}(t)$ is the envelope of
$V_{\textrm{inc}}(t)$ in direct analogy to
Eq.~\eqref{eq:envelope_statement}.

Equation~\eqref{eq:solved_I_time_equation} is a set of complex
first order linear differential equations analogous to
Shr\"{o}dinger's equation. By truncating the spectrum to a finite
number of modes, it is possible to solve
Eq.~\eqref{eq:solved_I_time_equation} numerically via standard
numerical integration techniques. In our case, we choose
forth-order Runga Kutta.  We generate the values of $k_n^2-k_0^2$
by generating 600x600 random matrices from the Gaussian Orthogonal
Ensemble, finding the spectrum, and unfolding it such that the
$k_n^2-k_0^2$ have a uniform density. We also generate the 600
$w_n$ as Gaussian random variables with 0 mean and width 1. All of
the remaining variables (including the initial conditions) are
physical parameters that must be set to match the situation we
wish to simulate.

For the runs displayed in this paper, we chose $R_R(\omega_0
c)/Z_0=1$, $\omega_0=22.5\textrm{ GHz}$, and $\Delta=10 \textrm{
m}^{-2}$. The $k_n$ were chosen to lie between $\approx 51\textrm{
m}^{-1}$ and $93\textrm{ m}^{-1}$. For initial conditions,
$V_n(0)=0$.  The envelope of the incident pulse,
$V_{i,\textrm{env}}$, had the form
\begin{equation}\label{eq:V_end_form}
    V_{i,\textrm{env}}(t)=e^{-(t \sigma_{\omega}-5)^2/2}
\end{equation}
with $\sigma_{\omega}=150 \textrm{ MHz}$.


\end{document}